\documentclass{ptapap}
\usepackage{amsmath}
\usepackage{amssymb}

\author{S. M. Rucinski}[UofT]
\affil[UofT]{Department of Astronomy and Astrophysics, University of Toronto, 
     60\ St.George Street, Toronto, ON, Canada M5S~3H4}

\title{Polish Astrophysics: \\The First Half-Century, 1923 - 1973}

\begin{document}

\maketitle

\begin{abstract}

An attempt is made to evaluate progress of the Polish astrophysical research of stars and of 
the inter-stellar medium (ISM) on the basis of scientific paper citations in the ADS 
database\footnote{This research has made use of NASA 
Astrophysics Data System Bibliographic Services.}.
Rather modest citation levels were observed in the years before the mid-1950's.
In the years 1958 -- 1973, thanks to the partly opened foreign contacts 
and to strong support from astronomers of the older generation,  
work of a number of young, energetic enthusiasts 
reached the world science levels and formed a strong basis for 
the well recognized, international successes of the next generations.
\end{abstract}

\section{Introduction}   

I feel honoured and very pleased being given an opportunity to deliver 
my talk in Toru\'n. This is happening 
exactly 60 years after I spent one month at the Piwnice Observatory 
learning the ropes in basic astronomy during my academic studies at the Warsaw University. 
My goal is not related to personal experiences, however: The task is
of evaluating progress of the Polish astrophysics since the formation of our Society in 1923.
It would be a very ambitious goal if intended to encompass the whole hundred
years of the Polish Astronomical Society.
While it is extremely encouraging and pleasing to witness the current, well-grounded  
position of the Polish astrophysics, it may be sufficiently interesting to look 
at the first half of those hundred years, the events-laden 
years 1923 -- 1973. The tool used will be publication citations as currently listed in the  
Astrophysics Data System (ADS) in the United States. 
The main advantage of the ADS records is that they form  
an objective measure of tangible contribution to the world's science. It is an unbiased
view and is open to access to anybody interested. It has been serving for many
purposes, among them 
as professional-progress performance indicators for job and grant applications.  
The ADS system is known for its high accuracy, ease of use and the
excellent, professional maintenance.

The term ''astrophysics'' was rarely used in the 1920's. 
A glance at the majority of papers of that time shows astronomy
still concentrated on stellar positions and motions of the Solar System bodies. 
Group spatial motions of stars are studied using newly-developed mathematical tools, 
but observational accuracy lags behind and sets severe 
limits on a more substantial progress. Studies of physical properties of celestial
bodies had then only started; it is only now that we recognize some research 
of that time as astrophysical in nature. 
In 1920, the ``Great Debate'' between Shapley and Curtis 
confronts the Big Question: Is the Milky Way -- with the already recognized off-centre solar position
-- the whole Universe? Or -- to the contrary -- perhaps,
the small, seemingly insignificant ''spiral nebulae'' at its periphery -- apparently
avoiding the Milky Way plane -- similar to it?
Soon, in October 1923, Edwin Hubble, using the 100-inch telescope located in an excellent climate,
finds that the largest of these "spiral nebulae", the one in Andromeda,  
has stars identical to those in the Milky Way. This discovery may count as the birth of the
stellar astrophysics; coincidently, it is the year when our Society starts its existence. 
 
The following half century, 1923 -- 1973, sees the progressively more extensive application of the
astrophysical work-horse, the spectrograph. It is also the time of revolutionary 
improvements in sensitivity and wavelength coverage of light detectors, first 
through rapidly improved photographic plate emulsions, 
then with the very sensitive photomultiplier --
the first device of the modern, electronic era. The two directions, 
the stellar astrophysics and the astrophysics of the inter-stellar matter (ISM) 
start dominating astronomy and  acquire the name of astrophysics.

\begin{figure}[h]     
\begin{center}
\includegraphics[width=0.55\textwidth]{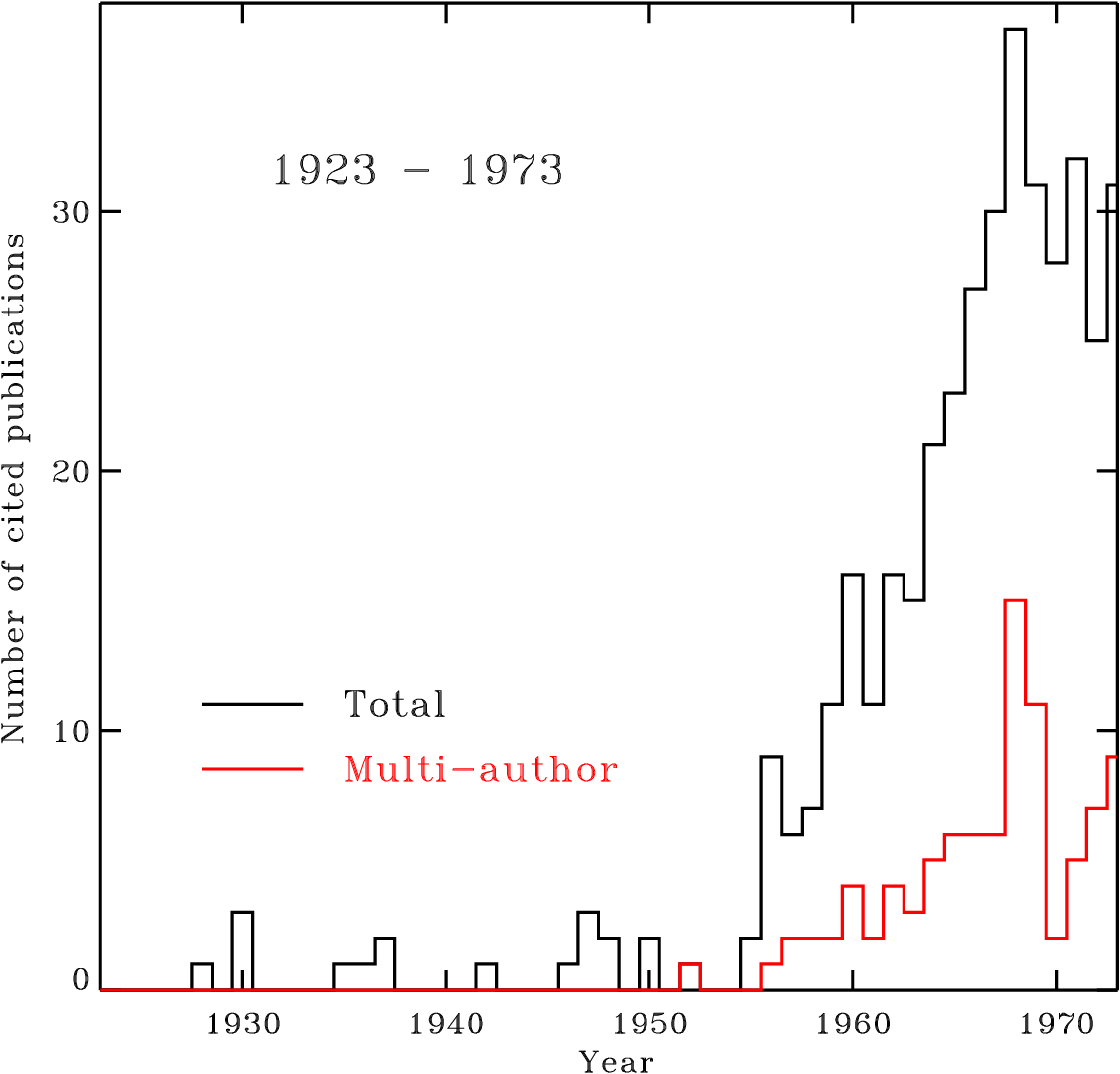}
\caption{The number of cited, multiple-author papers per year (the red line) 
   is compared with the total number of cited Polish publications (the black line).}
\label{fig:1}
\end{center}
\end{figure}      

\section{The citation data}

The present analysis has been done in the most straightforward way,
without any claims of professionalism in handling of
bibliographical data. It was started by forming a list of names of astronomers known to the author
as professionally active in the period 1923 -- 1973. The list was slightly expanded during the
lecture of already selected publications. 
An author on the list continued to be considered when any citations of 
his/her works appeared in the ADS, otherwise the name was dropped. 
Each paper was  checked for its content which had to be on astrophysics 
of stars or of the ISM. These two categories were considered separately with a possibility of
a minor overlap for stars observed using the polarimetric techniques; 
the object studied was used as the criterion for inclusion in either of the two categories.
The data noted 
were, for each author ($i$) and each paper ($j$):
(1)~the number of citations for the paper $c_{\rm ij}$ as of June 2023,
(2)~the number of co-authors ($k_{\rm ij}$) of that paper. 

The final total number of the author/citation entries was 452. 
Subsequently, the citations have been used in the normalized form, 
$c_{\rm ij}/k_{\rm ij}$; only such a form is a fair measure of the publication performance. 
The number of Polish astronomers contributing was 59. Some papers were
co-authored by foreign collaborators; then their presence  reduced the  normalized
count for the domestic authors. The total number of cited papers was 
396, in that as many as 303 were single-author publications. Thus, 
the relative contribution of the multi-authored papers was $23.5 \pm 2.6\%$
with the uncertainty estimated from the yearly statistics. 
We note that the dominance of the single-author
papers is entirely opposite to the current tendency of work in large teams. 
Figure~\ref{fig:1} shows the number of multi-author papers compared with the yearly totals.
As we will see later (Figure~\ref{fig:4} and the accompanying text), the 
author teams remained moderate in size throughout the whole period. 

\begin{figure}[h]     
\begin{center}
\includegraphics[width=0.65\textwidth]{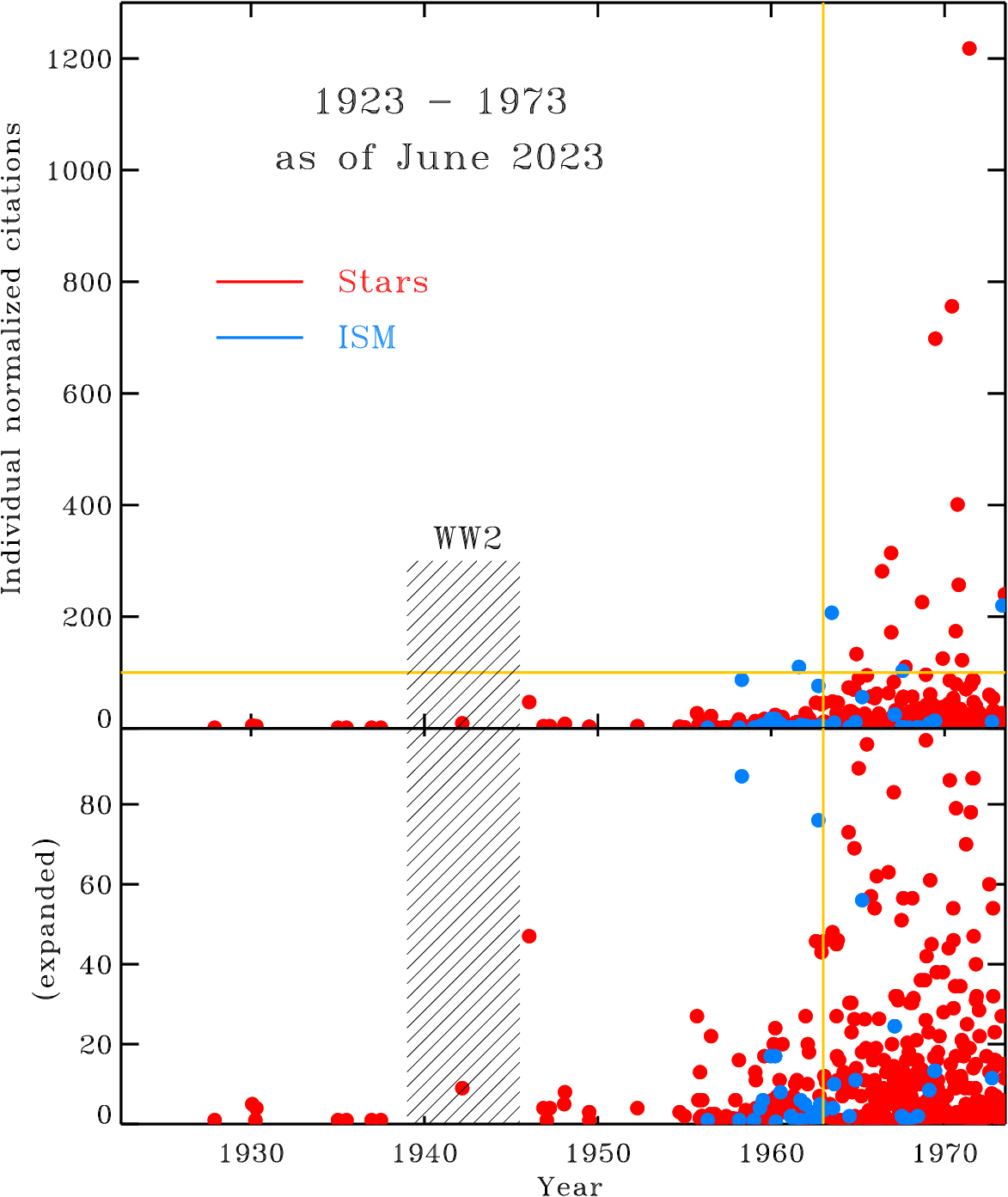}
\caption{The individual normalized citations for papers on stellar astrophysics (red dots) and 
     the ISM (blue dots). The greyed region shows the time of the second World War.
     The horizontal line delineates the region of
     normalized citations $<\!100$ shown in the expanded scale below. Similarly,
     the vertical orange line demarcates the last considered decade 1963 -- 1973.}
\label{fig:2}
\end{center}
\end{figure}      


Figure~\ref{fig:2} brings the essential statistics: Each point shows a normalized citation 
in years 1923 to 1973; the normalized citations below 100 are shown in the 
expanded scale in the lower part of the figure. 
The citation numbers are  low through
the whole considered period until the years 1956--1958: The early
Polish astrophysics is almost invisible in the world science. 
The first cited astrophysical paper is that by Stanis{\l}aw \citet{Szelig1926}  
in Wilno who worked on pulsating stars currently 
recognized as of the SX~Phe type; the material was
supplied by Dutch astronomers from Johannesburg in South Africa. 
The first spectroscopic paper   
was by Wilhelmina \citet{Iwan1936}
who worked in Stockholm and published the results in the
``Stockholm Annals'' -- one of several observatory publications existing at that time 
(we return to this subject later). 

Two other early papers may require a special attention.
The paper by Tadeusz \citet{Banach1942} carries the place and date of 
submission to the Astronomical Journal (AJ) of  ``Cracow 1941''. 
The degree of civilian-directed and intellectual-directed terror which was taking place 
at that time is well  documented and well known  to the inhabitants of Poland,
so it is hard to believe that the paper had been written and then
did reach the AJ Editors in 1942. 
It contains a description of an efficient use of the Cracovians -- a mathematical 
tool developed by Banachiewicz -- for solution of linear equations formed 
as an intermediate stage during least-squares parametric fits. It was not only a miracle 
that the paper appeared during that time, but that it also acquired as many as 9 citations,
a high number for a publication from before the mid-1950's.
Although written in  general terms, this paper of Banachiewicz 
played a role in early analyses of light-curves of eclipsing-binary stars; the
derivation via the ``Cracovian square root'' was much simpler and less prone to
mistakes in manual work than the standard Cramer's (determinant-based) approach.
(An aside comment: The Cracovians are definitely not differently-indexed 
matrices\footnote{https:/\!/www.impan.pl/\!zakopane/34/Koronski.pdf}.)

Another paper worth a note is an extensive work in ``Astrophysica Norvegica''
by Jeremi \citet{Wasiu1946}. It is actually a collection of
studies -- apparently forming his PhD Thesis -- on several aspects of stellar astrophysics.
It appeared in a book form;  parts of it may be still useful in some circumstances. 
The study had the largest number of citations (47) in the period before the 
large increase of the mid-1950's. 
Wasiuty\'{n}ski moved to Norway before the war and remained there retaining close
contacts with Poland. His study is surprisingly modern for the time; 
he was one of the first to consider the very new results 
on thermo-nuclear energy generation in stars and 
their impact on the models of stellar structure. 
He widely utilized power-law approximations and polytropic equations for the
stellar structure, a rather novel approach of that time.

We note that the impressive career of Bohdan Paczy\'nski started during the
years considered here. This period witnessed his first cited paper 
(with \citet{BeP1958}) currently at 1.33 normalized citations and then --
within a few years -- a dramatic increase of the citations reaching 756 and 
then 1218 citations for his two continually popular review papers, \citet{BeP1970}
and \citet{BeP1971}. Both were his single-author publications.

\section{The decade of the great success: 1963 -- 1973}

\begin{figure}[h]     
\begin{center}
\includegraphics[angle=90,width=0.75\textwidth]{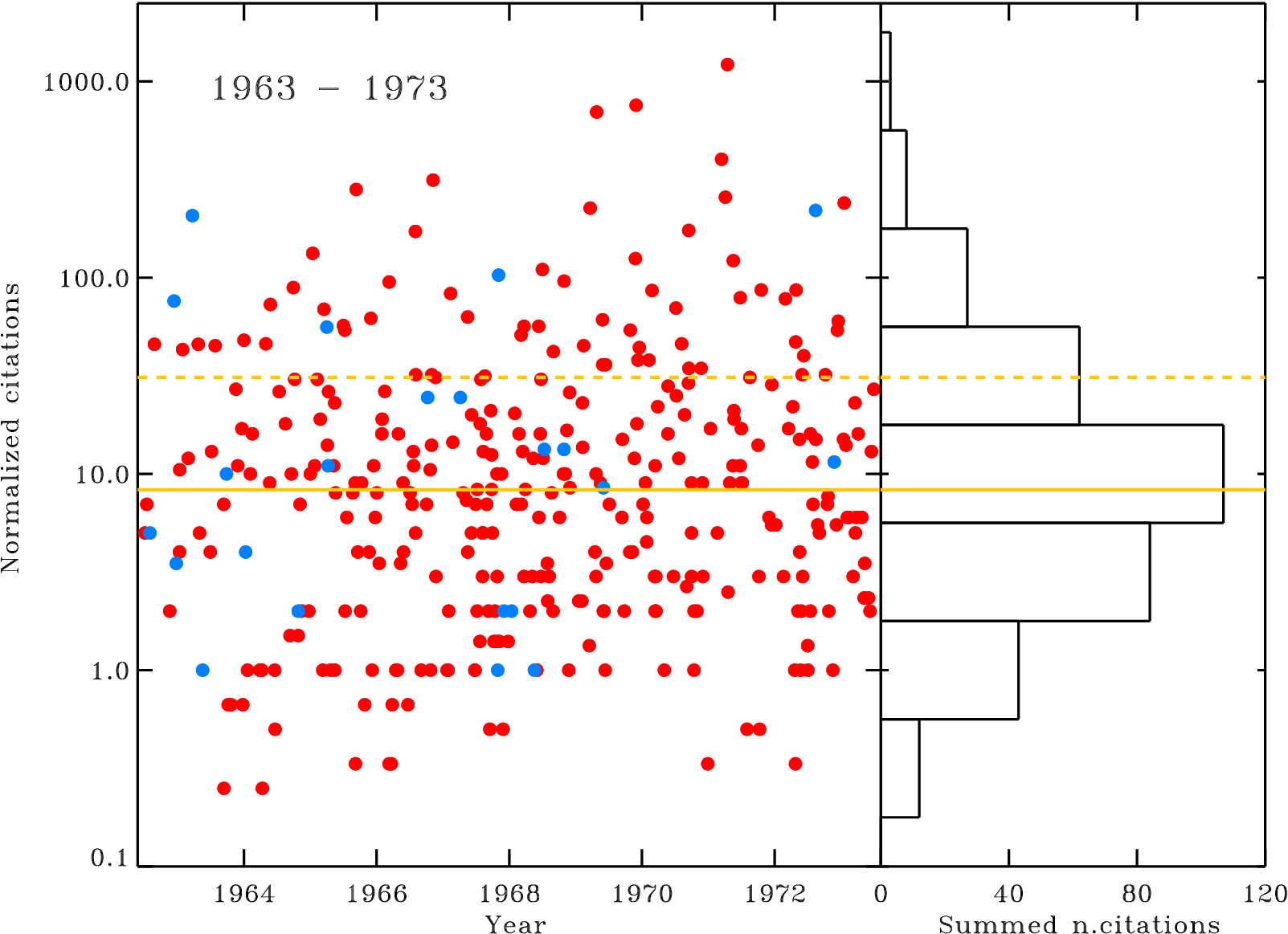}
\caption{The individual normalized citations for papers on  
     stellar astrophysics (red dots) and 
     the ISM (blue dots) for the last decade, 1963 - 1973. The data for that decade
     are the same as in Figure~\ref{fig:2} but  here shown in the logarithmic vertical scale. 
     The corresponding distribution of the 
     summed normalized citations is shown in the right panel.}
     \label{fig:3}
\end{center}
\end{figure}

A big change in the citation levels is noticeable in Figure~\ref{fig:2} 
within a few years around 1956 -- 1963. 
This was the time of important changes when scientific information started 
flowing through the formerly impenetrable political borders. 
Polish astronomers fully utilized these opportunities; a few were even able to travel and
work in Western observatories.  There was 
a lot to catch up: Astronomical instrumentation practically did not exist, 
destroyed or stolen by the invaders, with remnants of dubious use for astrophysics.  
The generation of pre-war-educated senior teachers was instrumental in creating
a new atmosphere where enthusiastic, energetic young scientists could 
start modern, astrophysical research. The accessible areas were: 
observationally -- photometry of a multitude of types of variable stars
and of the ISM, theoretically --  the internal structure of stars and
magnetohydro-gravito-dynamical instabilities  in the ISM.
The citation levels went up dramatically during the years 1956 -- 1963
from a few per year to tens and then hundreds per paper: 
Suddenly Polish astronomers became visible in the world. 
The transformation was amazingly fast so that -- by about 1963 --
the numbers of cited papers as well as normalized citations per paper 
reached world standards. As shown in Figure~\ref{fig:3}, the last 
decade is characterized by a broad distribution of generally high numbers of
citations. The modern Polish astrophysics started its existence.

\begin{figure}[h]     
\begin{center}
\includegraphics[width=0.55\textwidth]{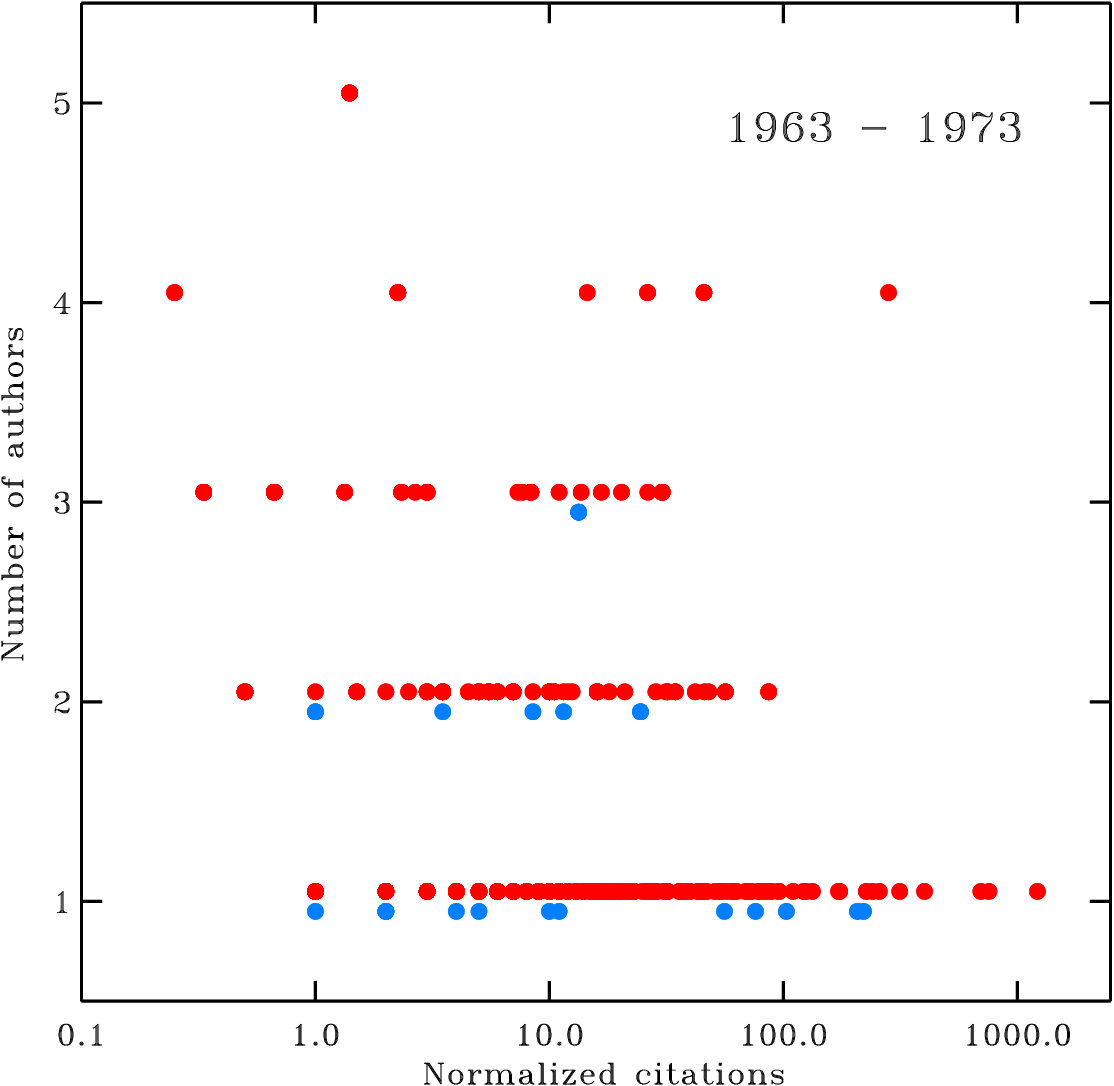}
\caption{The normalized citations for the last decade, 1963 -- 1973, are shown in the logarithmic 
     scale. They are split into groups depending on the number of co-authors (the vertical scale). 
     The division of symbols for
     the two groups of stellar (red) and ISM (blue) astrophysical papers is the same as
     used in previous figures. }
     \label{fig:4}
\end{center}
\end{figure}      

The  data for the decade 1963 - 1973, as shown in Figure~\ref{fig:3}, appear to indicate a 
very broad distribution of the normalized-citations 
characterized by a median of  8.5 and a mean value of 31, both calculated in linear units; 
the numbers are for the ensemble of the single- and multi-authored papers.
The distribution looks somewhat similar to a Gaussian, but 
obviously this is a gross simplification: Points at the citation level of exactly unity correspond
to single-author, singly-cited papers while the fractional normalized citations are for
poorly cited, multi-authored papers. The tail at $>\!500$ is defined by 
three single-author papers which even currently  keep on acquiring high citations.

Did the progressively more common team-work papers 
in the decade 1963 -- 1973 show any advantage of that new
style of work from the point of view of the normalized citations? An attempt to answer this
question is somewhat surprising by being... negative. This is demonstrated in Figure~\ref{fig:4}
where we see a gentle progression to {\it smaller\/} numbers for the increasing team size: 
The single-authored papers show the mean 
normalized citation of 40, those with 2 authors of 15, with 3 authors of 9.  
The new style of research was apparently only starting to be become popular.
We note that handling of large datasets was still difficult at that time
and computers -- which simplify the team work -- were still rather rarely used.
A first indication of ``citation-profitable'', multi-author study may be 
the case of the catalogue paper of Wies{\l}aw Wi\'{s}niewski who worked
in a team of four authors led by the famous photometrist H.L.\! \citet{Wisn1966}: 
The normalized citation of 281.25 (=1125/4) is still a very high number.

 \begin{figure}[h]     
\begin{center}
\includegraphics[width=0.55\textwidth]{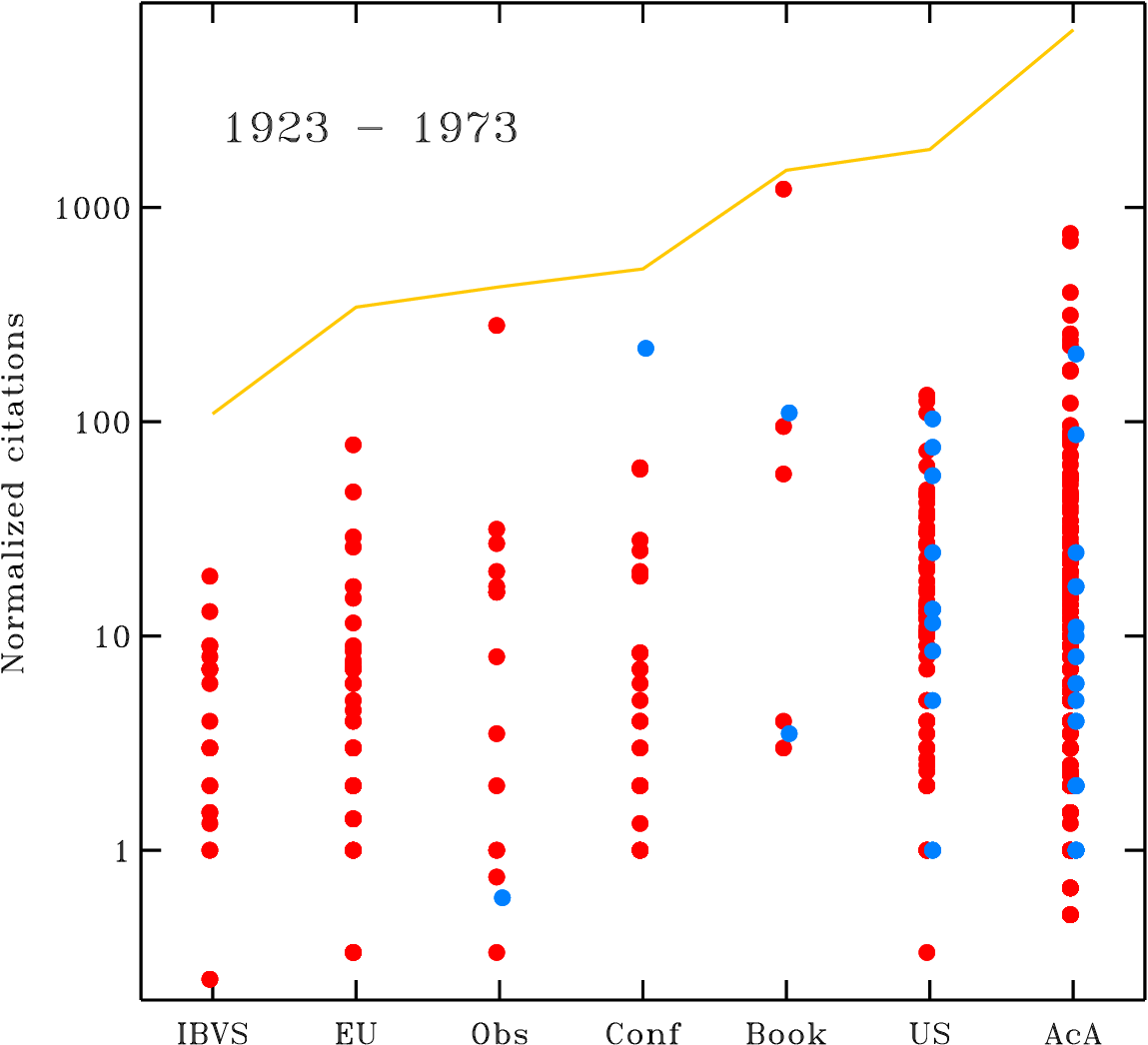}
\caption{The normalized citations for the half-century 1923 to 1973 -- with the
   coloured symbols as in the previous figures -- are divided into seven categories, 
   as described in the text. The orange line in the upper part of the figure gives 
   the summary normalized-citation count for each category. } 
   \label{fig:5}
\end{center}
\end{figure}      

\section{Where were the papers published?}

Polish astrophysical papers of the half-century 1923 -- 1973 appeared in many different 
publications. Virtually all (98.7\%) were written in English; only 
a few papers with modest citation numbers were written in German and Russian. 
 
An attempt to illustrate the wide range of types of publications is demonstrated  
in Figure~\ref{fig:5}. The publications were divided into seven categories
arranged by to the {\it summed normalized count\/} 
for a given category. Thus, at the right with the largest total count,
we see {\it Acta Astronomica\/} ({\bf AcA})
which apparently was the best publication vehicle for the Polish astronomical
community; it published papers acquiring citations in a wide range, 
from very small to impressive numbers approaching one thousand.
The publication, founded in 1925 by Banachiewicz has flourished after 
its move to Warsaw in 1967 and remains the very important publication tool in this country.

Citations resulting from the US publications ({\bf US}), 
 {\it The Astronomical Journal\/} and {\it The Astrophysical Journal\/}, 
 produced a large combined citation count. 
 This shows that the young, energetic astronomers were amazingly efficient in utilizing
the relatively short interval when US observatories were open to the Polish astronomers
of that time. 

The two next categories, the invited, in-depth articles in books ({\bf Book})
and the conference  reviews ({\bf Conf}) papers are somewhat similar in style.
The first category is practically by now extinct (the remaining publication is 
{\it The Annual Review of Astronomy and Astrophysics\/}), but several similar, 
by-invitation publications existed and competed at that time. 
Invited contributions by young Polish
astronomers \citep{Serk1962,Krusz1966,Smak1966,BeP1971} 
were a clear sign of their recognized prestigious position in the world science.

The observatory publications ({\bf Obs}) have all disappeared by now. They were related
to specific observatories and published large table or figure collections, all in view
of the currently difficult-to-imagine absence of any media able to convey large datasets:
All data had to be re-typed or re-plotted for any any subsequent use.
The  mentioned publications by \citet{Iwan1936} or Wi\'sniewski \citep{Wisn1966} 
belonged to this category. 

Independent national journals ({\bf EU}) were popular in Europe in the years 1923 -- 1973, 
but in the 1960's many struggled with lack of attractive material.  
Prompted by the creation of the European Southern Observatory in 1962,
several of them got amalgamated in 1968 into {\it Astronomy and Astrophysics\/}.
Among the already mentioned papers -- and  excluding those in {\it Acta Astronomica\/}
-- the papers by \citet{Szelig1926} or \citet{Wasiu1946} appeared in such by-now defunct 
publications.

Finally the {\it Information Bulletin of Variable Stars\/} ({\bf IBVS}) from Budapest. 
It is singled out here as a separate category because of its great popularity 
among Polish astronomers of that time; it was because a large amount of their 
research  concerned variable stars. The bulletin provided quickly released 
and -- in spite of a limited scope -- sometimes extensively cited publications.
The Bulletin ceased to exist in 2019. 

\section{Conclusions}

The available normalized-citations data  for the astrophysical papers
in the ADS system form an unbiased record of the large, important changes in the Polish
astronomy in the years 1923 -- 1973. After a long period of protracted difficulties, 
astrophysics of stars and of the inter-stellar medium (ISM) rapidly
developed in years 1956 -- 1963 reaching a high, world level in 
the last considered decade, 1963 -- 1973. The development was 
mainly due to the opening of the borders to the scientific information flow.
The first, bulky (``main-frame'') computers permitted revolutionary work on stellar
models. 
The dramatically improved sensitivity and accuracy of recently-introduced 
photomultipliers permitted observational studies of variable phenomena in stars and --
then novel -- polarimetric broad-band studies of the interstellar dust. 
We should note that lack of access to spectrographs on large telescopes 
still restricted detailed work on stellar atmospheres and on the gaseous phase of the ISM. 

The  presented work has a limitation: 
The inquiry of the ADS in June 2023 gives an instantaneous ``frozen-in'' 
view of the citations record. While this is useful as a first-look view,
it does not give any information on how citations varied in time. 
We see  some popular papers still acquiring new citations, 
while some papers stopped doing so long time ago. Is this a sign of a more 
general value {\it versus\/}  a short-lasting, but very timely contribution? A study of the 
citation time-profiles could be an interesting addition to the current analysis.

\acknowledgements{I would like to thank the Polish Astronomical Society, 
represented by the President Prof.\ Marek Sarna, for the invitation
to deliver the talk about beginnings of the Polish astrophysics and for 
the related travel support from the Society.}

\bibliographystyle{ptapap}
\bibliography{Rucinski}

\end{document}